\documentstyle[12pt]{article}
\begin{document}

\tolerance=5000
\def\pp{{\, \mid \hskip -1.5mm =}}
\def\cL{{\cal L}}
\def\be{\begin{equation}}
\def\ee{\end{equation}}
\def\bea{\begin{eqnarray}}
\def\eea{\end{eqnarray}}
\def\tr{{\rm tr}\, }
\def\nn{\nonumber \\}
\def\e{{\rm e}}
\def\D{{D \hskip -3mm /\,}}

\def\SEH{S_{\rm EH}}
\def\SGH{S_{\rm GH}}
\def\AdS5{{{\rm AdS}_5}}
\def\S4{{{\rm S}_4}}
\def\gfv{{g_{(5)}}}
\def\gfr{{g_{(4)}}}
\def\SC{{S_{\rm C}}}
\def\RH{{R_{\rm H}}}

\def\wlBox{\mbox{
\raisebox{0.1cm}{$\widetilde{\mbox{\raisebox{-0.1cm}\fbox{\ }}}$}}}
\def\htBox{\mbox{
\raisebox{0.1cm}{$\hat{\mbox{\raisebox{-0.1cm}{$\Box$}}}$}}}

\ 

\vskip -3cm

\  \hfill
\begin{minipage}{2.5cm}
October 2001 \\
\end{minipage}

\vfill

\begin{center}
{\large\bf 
De Sitter space versus Nariai Black Hole: stability in d5 higher derivative 
gravity.
}

\vfill

{\sc Shin'ichi NOJIRI}\footnote{nojiri@cc.nda.ac.jp} 
and {\sc Sergei D. ODINTSOV}$^{\spadesuit}$\footnote{
On leave from Tomsk State Pedagogical University, 
634041 Tomsk, RUSSIA. \\
odintsov@ifug5.ugto.mx, odintsov@mail.tomsknet.ru} \\

\vfill

{\sl Department of Applied Physics \\
National Defence Academy,
Hashirimizu Yokosuka 239-8686, JAPAN}

\vfill

{\sl $\spadesuit$
Instituto de Fisica de la Universidad de Guanajuato, \\
Lomas del Bosque 103, Apdo. Postal E-143, 
37150 Leon,Gto., MEXICO}

\vfill

{\bf ABSTRACT}

\end{center}

d5 higher derivative gravity on the Schwarzschild-de Sitter (SdS) black hole 
background is considered. Two horizons SdS BHs are not 
in thermal equilibrium and Hawking-Page phase transitions are not 
expected there, unlike to the case of AdS BHs. It is demonstrated that 
there exists the regime of d5 theory where Nariai BH which is extremal 
limit of SdS BH is stable. It is in the contrast with Einstein gravity on 
such background
where only pure de Sitter space is always stable. Speculating on the 
applications in proposed dS/CFT correspondence, these two (de Sitter and 
Nariai) stable spaces may correspond to confining-deconfining phases 
in dual CFT. 

\vfill


\newpage

The increasing evidence indicating a positive cosmological constant for 
our universe calls to better study of de Sitter gravity. One of the 
fundamental questions in de Sitter gravity is related with holographic 
principle which is presumably realized there in the form of dS/CFT 
correspondence suggested in ref.\cite{strominger} (for earlier attempts 
on dS/CFT duality see \cite{earlier}). Despite the number of efforts 
\cite{ds/cft,r2} (for a recent review, see \cite{lecture}) still dS/CFT 
correspondence is not understood on the same level as AdS/CFT 
correspondence. Moreover, the corresponding consistent dual CFT is  
not yet formulated. Nevertheless, the hunt for dS/CFT correspondence 
continues.
It is expected that analogy with AdS/CFT may be often helpful in such 
investigation.

It has been observed quite long ago by Hawking and Page \cite{HP} that 
AdS black hole (BH) thermodynamics admits phase transitions in the following 
way: low temperature BHs are not stable and they decay into global AdS 
space. High temperature BHs are more stable than global AdS space.
Hawking-Page phase transitions for 5d AdS BHs are very important in 
frames of AdS/CFT.
Indeed, they were interpreted by Witten \cite{witten} as corresponding 
to a deconfinement-confinement transition in the large-$N$ limit of 
an ${\cal N}=4$ $SU(N)$ super Yang-Mills theory living on the boundary 
of 5d AdS BH.

The interesting question appears: Can the similar effect be expected for 
proposed dS/CFT correspondence? From first look it seems that the answer 
is completely negative. The reason is that 
there are two Hawking temperatures in Schwarzschild-de Sitter (SdS) BH
(which is the natural analog of AdS BH) since 
there are two horizons, black hole one and cosmological one. 
Then SdS spacetime is not in the thermal equilibrium unlike to AdS BH.
Moreover, it is known that pure de Sitter space is the only stable space
among SdS backgrounds for Einstein gravity.  
There is, however, second exception. It is  the Nariai space, 
which is the extremal limit where the black hole horizon coincides 
with the cosmological one. 

In the present Letter we discuss the question of stability of dS and 
Nariai BH in higher derivative gravity (for general introduction to such
theory, see 
\cite{BOS}). We will show that despite the absence of Hawking-Page phase 
transitions, there is some theory regime (defined by coefficients of 
higher derivative terms) 
where Nariai BH is stable and does not decay into pure de Sitter space.
Brief speculation of the relevance of this observation to proposed dS/CFT 
correspondence is made.

The general action of $d+1$ dimensional $R^2$-gravity is given by 
\be
\label{Svi}
S=\int d^{d+1} x \sqrt{-\hat G}\left\{a \hat R^2 
+ b \hat R_{\mu\nu} \hat R^{\mu\nu}
+ c  \hat R_{\mu\nu\xi\sigma} \hat R^{\mu\nu\xi\sigma}
+ {1 \over \kappa^2} \hat R - \Lambda\right\} \ .
\ee
For simplicity, we only consider the $c=0$ case, for a while. 
Then Schwarzschild-anti de Sitter or 
Schwarzschild-de Sitter spacetime is an exact solution:
\bea
\label{SAdS}
&& ds^2=\hat G_{\mu\nu}dx^\mu dx^\nu 
=-\e^{2\rho_0}dt^2 + \e^{-2\rho_0}dr^2 
+ r^2\sum_{i,j}^{d-1} g_{ij}dx^i dx^j\ ,\nn
&& \e^{2\rho_0}={1 \over r^{d-2}}\left(-\mu + {kr^{d-2} \over d-2} 
+ {r^d \over L}\right)\ .
\eea
Here $g_{ij}$ is the metric of the $(d-1)$-dimensional 
Einstein manifold, which is 
defined by $R_{ij}=kg_{ij}$, where $R_{ij}$ is the Ricci 
tensor defined by $g_{ij}$ and $k$ is a constant. 
For example, we have $k>0$ for the sphere, $k<0$ for the 
hyperboloid, and as a special case, flat space for $k=0$. 
The parameter $L$ in (\ref{SAdS}) is related with the length 
parameter of the asymptotic AdS or dS spacetime and is found  
solving the equation
\be
\label{ll}
0={d^2(d+1)(d-3) a \over L^2} + {d^2(d-3) b \over L^2} 
- {d(d-1) \over \kappa^2 L}-\Lambda\ .
\ee
If $L>0$, the spacetime is SAdS and if $L<0$, SdS. Here we 
consider the case of $L<0$ and $d=4$. In dS, we cannot 
embedd the hyperbolic or the flat space as the surface with 
constant $r$. Then one only discusses the case that $g_{ij}$ 
 (\ref{SAdS}) is the metric of the unit sphere
\be
\label{us}
\sum_{i,j}^3 g_{ij}dx^i dx^j = d\Omega_3^2\ .
\ee
In this case $k=2$. Defining the length parameter $l$ by 
\be
\label{Ni}
l^2 = -L\ , 
\ee
$\e^{2\rho}$ in the metric (\ref{SAdS}) has the 
following form
\be
\label{Nii}
\e^{2\rho_0}={1 \over r^{d-2}}\left(-\mu + r^2 
- {r^4 \over l^2}\right)\ .
\ee
Then there are two horizons, where $\e^{2\rho_0}=0$, at
\be
\label{Niii}
r=r_{c,bh}^2\equiv {l^2 \pm \sqrt{l^4 - 4\mu l^2} \over 2}\ ,
\ee
where the plus sign corresponds to the cosmological horizon 
$r=r_c$ and the minus to the black hole one $r=r_{bh}$. 
The  corresponding Hawking temperature $T_H$ is given by
\be
\label{Niv}
T_H={1 \over 4\pi}\left|\left.{d\left(\e^{2\rho_0}\right) \over 
dr}\right|_{r=r_{c,bh}}\right|
= {1 \over 2\pi}\left|{\mu \over r_{c,bh}^3} - {r_{c,bh} 
\over l^2}\right|\ .
\ee
For the pure dS case, where $\mu=0$, we find
\be
\label{Nv}
r_{bh}=0\ ,\quad r_c=l\ ,\quad T_H={1 \over 2\pi l}
\ee
and for the Nariai space \cite{N}, where $\mu={l^2 \over 4}$,
\be
\label{Nvi}
r_{bh}=r_c={l \over \sqrt{2}}\ , \quad T_H={15 \over 
2\sqrt{2}l\pi}\ .
\ee
We should note that when $c\neq 0$ in (\ref{Svi}), the 
general Schwarzschild-(anti) de Sitter spacetime is not the 
exact solution. Nevertheless,  even if $c\neq 0$, the pure dS and 
the Nariai space are exact solutions since the Riemann 
curvature is covariantly constant:
\be
\label{Riemann}
\hat{R}_{\mu\nu\xi\sigma} 
= {1\over l^2}\left( \hat{G}_{\mu\xi}\hat{G}
_{\nu\sigma} -\hat{G}_{\mu\sigma}\hat{G}_{\nu\xi} \right). 
\ee
Here, instead of (\ref{ll}), the length parameter $l^2$ 
is given by
\bea
\label{ll2}
0&=&{a \over l^4}(d+1)d^2(d-3) + {b \over l^4}d^2(d-3) \nn
&& + {2c \over l^4}d(d-3) + {d(d-1) \over \kappa^2 l^2}
-\Lambda\ .
\eea
Then in what follows we also consider the case of $c\neq 0$.

Before going forward, we give some remarks about the Nariai 
space, which is given in the Nariai limit 
$\mu\rightarrow {l^2 \over 4}$. Before taking the limit, one changes 
the coordinate $(t,r)$ to $(\tau,\theta)$ by
\be
\label{Nvii}
t={l \over \sqrt{l^2 - 4\mu}}\tau\ ,\quad 
r^2 = {l^2 \over 2}- \cos\theta {\sqrt{l^4 - 4\mu l^2} 
\over 2}\ .
\ee
Then the black hole horizon corresponds to $\theta=0$ and 
the cosmological one to $\theta=\pi$. In the coordinates 
$(\tau,\theta)$, the metric of SdS is rewritten by
\be
\label{SdS}
ds^2 = - {\sin^2 \theta \over 
2\left(l^2 - \cos\theta \sqrt{l^4 - 4\mu l^2}\right)}d\tau^2
+ {l^2 \over 2}d\theta^2 
+ \left({l^2 \over 2} - \cos\theta \sqrt{l^4 - 4\mu l^2} 
\right)d\Omega_3^2\ .
\ee
Then by taking the Nariai limit $\mu\rightarrow {l^2 \over 4}$, 
one finds
\be
\label{Nr}
ds_{\rm Nariai}^2 = - {\sin^2 \theta \over 
2}d\tau^2 + {l^2 \over 2}d\theta^2 
+ {l^2 \over 2} d\Omega_3^2\ .
\ee
If we Wick-rotate the time coordinate $\tau$ by
\be
\label{Nviii}
\tau\rightarrow il\tilde \tau\ ,
\ee
the metric (\ref{Nr}) is the direct product of S$_2$ and $S_3$ 
with the radius ${l \over \sqrt{2}}$ and we find $\tilde \tau$ 
should have the periodicity of $2\pi$. 

In 4 dimensional ($d=3$) Nariai space, there occurs a 
very interesting 
phenomenon called ``anti-evaporation'', which was first 
found in \cite{BH} by using 2d trace anomaly induced 
effective action including dilaton \cite{BHano,NOano,A}.
It corresponds to quantum expansion of BH. (It is not yet completely clear
if this is fundamental or just transitionary effect).  
This phenomenon has been confirmed by using 4d trace anomaly  
induced effective action \cite{NOnari}.

Let us discuss the free energies of the pure dS and the Nariai 
space. Since for these cases, the scalar, Ricci and Riemann 
curvatures are given by
\be
\label{rr}
\hat{R}= {20 \over l^2}\ ,\quad \hat{R}_{\mu\nu}=
{4\over l^{2}}G_{\mu\nu}\ ,\quad \hat{R}_{\mu\nu\xi\sigma} 
= {1\over l^2}\left( \hat{G}_{\mu\xi}\hat{G}
_{\nu\sigma} -\hat{G}_{\mu\sigma}\hat{G}_{\nu\xi} \right)\ . 
\ee
the action (\ref{Svi}) with $d=4$ is given by
\bea
\label{Nix}
S&=&\left({400a \over l^4}
+ {80 b \over l^4} + {40 c \over l^4}
+ {20 \over \kappa^2} - \Lambda\right)
\int d^5 x \sqrt{-\hat G} \nn
&=&\left({320a \over l^4}
+ {64 b \over l^4} + {32 c \over l^4}
+ {8 \over l^2\kappa^2} \right)
\int d^5 x \sqrt{\hat G} \ .
\eea
In the second line of (\ref{Nix}), Eq. (\ref{ll}) is used. 
For pure de Sitter space, one gets
\be
\label{Nx}
V_{\rm dS}=\int d^5 x \sqrt{\hat G}={V_3 \over T_H}\int_0^{r_c=l} dr 
r^3 = {V_3l^4 \over 4 T_H}
={V_3 \over 4T_H\left(2\pi T_H\right)^4}\ ,
\ee
and for the Nariai space
\be
\label{Nxi}
V_{\rm Nariai}=\int d^5 x \sqrt{\hat G}=V_2V_3 \left({l^2 \over 2}
\right)^{5 \over 2}={8\cdot 15^5 \over \left(2\pi
\right)^{5 \over 2}}{V_3 \over 4T_H\left(2\pi T_H\right)^4}\ ,
\ee
Here $V_2=4\pi$ and $V_3$ are the volumes of  2 sphere and 
3 sphere, respectively. 
In (\ref{Nxi}) it is assumed $\tilde\tau$ in (\ref{Nviii}) has 
the period $2\pi$. As it follows from (\ref{Nvii}) and 
(\ref{Nviii})
\be
\label{Nxib}
i\tilde\tau = {t\sqrt{l^2 - 4\mu} \over l^2}\ .
\ee
Then if $t$ has the periodicity of ${1 \over T_H}$ (near the 
Nariai limit) when 
Wick-rotated, the periodicity $\tilde P$ of $\tilde\tau$ 
should be  
\be
\label{Nxic}
\tilde P = {\sqrt{l^2 - 4\mu} \over T_H l^2}\ .
\ee
Then the volume $V_{\rm Nariai}$ of the Nariai space in 
(\ref{Nxi}) should be modified as 
\be
\label{Nxid}
V_{\rm Nariai}\rightarrow {\tilde P \over 2\pi}V_{\rm Nariai}\ ,
\ee
which, we should note, vanishes in the Nariai limit 
$\mu\rightarrow {l^2 \over 4}$ since $\tilde P$ vanishes. 

Using (\ref{Nix}) and (\ref{Nx}), we find that 
the free energy $F=-T_H S$ for the pure dS is given by
\be
\label{Nxii}
F_{\rm dS}=-\left({320a \over l^4}
+ {64 b \over l^4} + {32 c \over l^4} 
+ {8 \over l^2\kappa^2} \right)
{V_3 \over 4\left(2\pi T_H\right)^4}\ ,
\ee
 Using (\ref{Nix}), (\ref{Nxi}) and (\ref{Nxid}), 
 one gets the free energy of the Nariai space: 
\be
\label{Nxiii}
F_{\rm Nariai}=-\left({320a \over l^4}
+ {64 b \over l^4}  + {32 c \over l^4} 
+ {8 \over l^2\kappa^2} \right)
{8\cdot 15^5 \over \left(2\pi
\right)^{5 \over 2}}
{V_3 \over 4\left(2\pi T_H\right)^4}{\tilde P \over 2\pi}
\stackrel{\mu\rightarrow {l^2 \over 4}}{\longrightarrow} 0\ .
\ee
Therefore the pure dS is more stable than 
the Nariai space if 
\be
\label{Nxv}
{320a \over l^4}
+ {64 b \over l^4}  + {32 c \over l^4} 
+ {8 \over l^2\kappa^2}>0\ ,
\ee
but the Nariai space becomes stable if
\be
\label{Nxvi}
{320a \over l^4}
+ {64 b \over l^4}  + {32 c \over l^4} 
+ {8 \over l^2\kappa^2}<0\ .
\ee
Therefore there is a critical point (or surface) at
\be
\label{Nxvii}
{320a \over l^4}
+ {64 b \over l^4}  + {32 c \over l^4} 
+ {8 \over l^2\kappa^2}=0\ .
\ee
It is easily seen that for pure Einstein gravity, the above equation 
has no solution. Hence, dS space is always stable there!

The expressions of the free energies (\ref{Nxii}) and 
(\ref{Nxiii})  seem to be strange since the fourth 
inverse power of the temperature appears in the expressions. 
As we are considering the limits, Nariai limit and the vanishing 
mass limit, the temperature does not depend on the black hole mass 
but only depends on the length parameter $l$ as in (\ref{Nv}) and 
(\ref{Nvi}). The parameter $l$ is not the dynamical 
variable. Of course, when we include the scalar fields $\phi_i$ 
with potential $V(\phi_i)$, the length parameter can be regarded 
as a dynamical variable by replacing the cosmological constant 
$\Lambda$ in the action (\ref{Svi}) by the potential 
$\Lambda\rightarrow V(\phi_i)$. One can also consider the 
condensation of the anti-symmetric tensor fields as in 
the usual AdS$_5$/CFT$_4$ scenario. If $l$ is not the dynamical 
variable, we should consider the radius of the black hole as a 
dynamical variable but when the radius changes from one of the 
(Nariai and the vanishing mass) limits, the system is not in 
the thermal equilibrium. Then one cannot define the heat 
capacity and (or) to use the other thermodynamical stability conditions 
as those developed in ref.\cite{CG}. This 
 makes difficult to argue about the 
local thermal stability. One can conjecture  that  two limits are 
connected with each other by some thermal inequilibrium process. 

  From the viewpoint of the WKB approximation of the path 
integral, the partition function $Z$ can be given by the 
classical action $S_{\rm cl}$, where the classical 
solution is substitued into the action (\ref{Svi}), 
\be
\label{Nxviii}
Z\sim \e^{S_{\rm cl}}\ .
\ee
The expression (\ref{Nxiii}) is valid even if the system is 
not in the thermal equilibrium. For the general SdS spacetime, 
one gets
\be
\label{Nxix}
V_{\rm SdS}=\int d^5 x \sqrt{-\hat G}
={V_3 \over T_H}\int_{r_{bh}}^{r_c} dr 
r^3 = {V_3l^4 \over 4 T_H}\sqrt{1 - {4\mu \over l^2}}\ .
\ee
Here we assume the time variable $t$ has a period 
${1 \over T_H}$ although the system is in the thermal 
inequilibrium. When $c=0$, the action $S_{\rm SdS}$ for SdS 
is also given by (\ref{Nix})
\be
\label{NixB}
S_{\rm SdS}=\left({320a \over l^4}
+ {64 b \over l^4} + {8 \over l^2\kappa^2} \right)
V_{\rm SdS}\ .
\ee
The actions for the pure dS and Nariai space are just given 
by replacing $V_{\rm SdS}$ with $V_{\rm dS}$ in (\ref{Nx}) 
and ${\tilde P \over 2\pi}V_{\rm Nariai}$ in (\ref{Nxi}) 
and (\ref{Nxid}), respectively. 
Since $\sqrt{1 - {4\mu \over l^2}}<1$, we 
find
\be
\label{Nxx}
V_{\rm dS}>V_{\rm SdS}>{\tilde P \over 2\pi}V_{\rm Nariai}=0\ .
\ee
Then the classical action for dS is larger than that for SdS 
if the condition (\ref{Nxv}) is satisfied. Therefore dS 
is stable even locally. On the other hand, if the 
condition (\ref{Nxvi}) is satisfied, the classical action 
for the Nariai space is larger than that for SdS. Then 
the Nariai space becomes stable even locally. 
Hence, we demonstrated that there is regime of d5 higher derivative gravity
where Nariai BH does not decay into pure dS space.

In case of AdS$_5$/CFT$_4$ correspondence higher derivative 
terms like $R^2$-terms appear as next-to-leading, ${1 \over N}$ correction. 
The ${\cal N}=2$ theory with the gauge group $Sp(N)$ arises as 
the low-energy theory on the world volume on $N$ D3-branes 
sitting inside 8 D7-branes at an O7-plane \cite{Sen}. 
The string theory dual to this theory has been conjectured 
to be type IIB string theory on ${\rm AdS}_5\times {\rm X}^5$ where 
${\rm X}_5={\rm S}^5/Z_2$ \cite{FS}, whose low energy effective 
action is given by  
\be
\label{bng3} 
S=\int_{{\rm AdS}_5} d^5x \sqrt{G}\left\{{N^2 \over 4\pi^2}
R-\Lambda 
+ {6N \over 24\cdot 16\pi^2}R_{\mu\nu\rho\sigma}
R^{\mu\nu\rho\sigma}\right\}\ .
\ee
Then $R^2$-term appears as $1/N$ correction. 
We should note that the coupling constants are chosen to be 
dimensionless by the proper redefinitions. Then one can identify 
\be
\label{bng4}
{1 \over \kappa^2}={N^2 \over 4\pi^2}\ ,\quad 
c={6N \over 24\cdot 16\pi^2}\ ,
\ee
In the model (\ref{bng}), $\Lambda$ is negative and the 
spacetime is always asymptotically AdS.
 
Here as a toy model, we consider the case 
that $\Lambda$ is positive and is given by
\be
\label{bng5}
\Lambda={12N^2 \over 4\pi^2} - {6N \over 24\cdot 16\pi^2}\ ,
\ee
Suppose that $c$ is negative:
\be
\label{bng6}
{1 \over \kappa^2}={N^2 \over 4\pi^2}\ ,\quad 
c=-{6N \over 24\cdot 16\pi^2}\ ,
\ee
It has been demonstrated in ref.\cite{r2} that in frames of dS/CFT 
correspondence such de Sitter higher derivative gravity reproduces 
the holographic conformal anomaly for above $Sp(N)$ super Yang-Mills 
theory. 
Then Eq.(\ref{ll2}) tells $l^2=1$ and Eq.(\ref{Nxvii}) gives 
the critical point at 
\be
\label{bng}
N={1 \over 4}\ 
\ee
Then from Eq.(\ref{Nxv}), the pure de Sitter space is stable 
when $N\geq 1$. Of course, since this model 
is not realistic dual CFT model one can expect that there will be and 
dS and Nariai phases when realistic dual CFT will be proposed. 
(Moreover, even in above model the next powers of the curvatures 
may qualitatively change the situation).
Then presumably one of these two phases will correspond to confinement,
while another one to deconfinement in dual CFT. In any case, the fact that
Nariai BH may be preferrable vacuum state for some dS gravitational theory
looks quite attractive. 

\ 

\noindent
{\bf Acknoweledgements.} We thank M. Cveti\v c and V. Tkach 
for helpful discussion. 
The research by SDO has been supported in part by 
CONACyT (CP, Ref.990356).

\end{document}